\documentclass[a4paper]{manuscript}

\usepackage[T1]{fontenc}

\usepackage{amsthm}
\usepackage{dsfont}
\usepackage{caption}
\usepackage{subcaption}

\theoremstyle{definition}
\newtheorem{definition}{Definition}[section]
\theoremstyle{remark}
\newtheorem{remark}{Remark}[section]
\theoremstyle{proposition}
\newtheorem{proposition}{Proposition}[section]
\theoremstyle{theorem}
\newtheorem{theorem}{Theorem}[section]
\newtheorem{corollary}{Corollary}[section]
\theoremstyle{assumption}
\newtheorem{assumption}{Assumption}[section]

\title{Non-coercive extortion in game theory}
\shorttitle{Non-coercive extortion}

\author[1,2,*]{María Alejandra Ramírez}
\author[3]{Rosemarie Nagel} 
\author[4,5,6,7]{David Wolpert}
\author[1,4,8,9]{J\"urgen Jost}

\affiliation[1]{\small{Max Planck Institute for Mathematics in the Sciences, Leipzig, Germany}}

\affiliation[2]{Max Planck Institute for Evolutionary Biology, Plön, Germany}

\affiliation[3]{\small{ICREA, Barcelona School of Economics, Universitat Pompeu Fabra, Barcelona, Spain}}

\affiliation[4]{Santa Fe Institute for the Sciences of Complexity, Santa Fe, NM, USA}

\affiliation[5]{International Center for Theoretical Physics, Trieste, Italy}

\affiliation[6]{Complexity Science Hub, Vienna, Austria}

\affiliation[7]{Arizona State University, Tempe, AZ, USA}

\affiliation[8]{Center for Scalable Data Analytics and Artificial Intelligence, Leipzig, Germany}

\affiliation[9]{Max Planck Institute for Human Cognitive and Brain Sciences, Leipzig, Germany}

\affiliation[*]{\texttt{ramirez@mis.mpg.de}}

\abstract{Commitments play a crucial role in game theory, shaping strategic interactions by either altering a player's own payoffs or influencing the incentives of others through outcome-contingent payments. While most research has focused on using commitments to achieve efficient equilibria, their potential applications beyond this goal remain largely unexplored. In this study, we introduce a non-coercive extortion mechanism that leverages commitments to outcome-contingent payments, demonstrating how a player or external agent can extract profit by offering rewards rather than threatening punishment. At the core of the mechanism is the introduction of sequentiality into a simultaneous-move game, fundamentally reshaping the strategic interaction. We derive the conditions under which extortion is successful, identify the class of games susceptible to this scheme, and determine both the maximum extractable profit and the minimum required payment. To illustrate the extortion mechanism, we apply it to 2×2 games, highlighting how even simple strategic settings can be vulnerable to this form of manipulation. Our results reveal strategic vulnerabilities in competitive settings, with significant implications for economic markets, diplomatic relations, and multi-agent systems operating in blockchain environments. This work broadens our understanding of commitments in game theory and raises critical questions about how to safeguard strategic interactions from exploitation through non-coercive extortion.\newline}

\keywords{\hspace{3.0mm}Non-coercive extortion | Commitments | Outcome-contingent payments | Strategic manipulation}

\begin{document}

\maketitle


Game theory provides a framework for analysing strategic interactions, where players make decisions to maximize their outcomes based on the actions of others \cite{vonNeumann:1947:book,maschler:2020:book,gonzalez:2023:book}. Within this framework, commitments are powerful tools that influence decision-making by altering the structure of these interactions. Since Thomas Schelling's seminal work \textit{The Strategy of Conflict} \cite{schelling:1960:book}, their impact has been explored across various fields such as economics, political science, and computer science. At their core, commitments enable players to reshape strategic decisions by modifying the underlying game describing the interaction.

Previous work has focused on how players can modify their own elements of the game, namely their own payoff functions or strategy sets. For instance, players may alter their own payoff functions by misreporting utilities \cite{crawford:1979:EcoLetters,sobel:1981:Econometrica}, burning money \cite{fudenberg:1991:book}, or by delegating decision-making authority to an agent with different incentives. This approach, central to the principal-agent literature, serves as a widely used commitment mechanism to influence competitive outcomes \cite{vickers:1985:EJ, fershtman:1987:AER,fershtman:1991:IER,katz:1991:RAND}. 
Alternatively, players can modify their strategy sets by committing to specific strategies before playing the game, as in Stackelberg scenarios. In these cases, a ``leader" moves first, committing to a course of action, while ``followers" observe this commitment and adjust their responses accordingly \cite{stackelberg:1934:book}. The implications of this order of play has been extensively analysed, especially in relation to first-mover advantages and how early commitments to strategy subsets can influence the strategic interaction \cite{hamilton:1990:GEB,vanDamme:1996:GEB,vanDamme:1999:GEB,harrenstein:2007:AMAS,renou:2009:GEB}.

In recent decades, there has been increasing interest in understanding how players' commitments influence the game structure of their co-players. A key focus has been on the ability to reshape a co-player's payoffs, particularly through outcome-contingent side payments, i.e. transfers from one player to another based on the co-player's chosen action \cite{jackson:2005:RES}. The main objective of these studies has been to explore the implementation of efficient equilibria \cite{prat:2003:Econometrica,calvo:2010:JEEA}, in line with Coase's theorem \cite{coase:1960:JLE}. In particular, the Prisoner's Dilemma, known for its Pareto-inefficient Nash equilibrium, has been a central case for examining how side-payments can enhance efficiency \cite{varian:1994:AER,andreoni:1999:PNAS,charness:2007:GEB,fong:2009:GEB}.

Although outcome-contingent payments in game theory have been largely studied in the context of achieving efficient equilibria, their potential applications beyond this goal remain largely unexplored. Here, we examine their role in extortion, showing how players and external agents can use them to manipulate strategic interactions. In particular, we formulate an extortion scheme that leverages commitments of outcome-contingent payments to extract a profit from players. Extortion is typically associated with coercion, where an individual is threatened with punishment unless they comply, allowing the extortioner to extract a profit \cite{ellsberg:1959:lecture}. In contrast, our scheme takes a counterintuitive approach: instead of punishment, the threat takes the form of a reward.
This non-coercive extortion mechanism has significant implications for strategic interactions between competitive players, including individuals in the marketplace, organizations, and nation-states. 
Beyond traditional human interactions, a new and growing application of this scheme lies in modern multi-agent systems, where artificial agents strategically interact with both humans and each other in real time \cite{rahwan:2019:Nature}. In such systems, emerging commitment mechanisms —such as smart contracts on blockchains— allow agents to make credible commitments, which can, in turn, facilitate extortion \cite{dafoe:2020:arXiv,hammond:2025:techrep}.

While prior research on extortion in game theory has mainly focused on employing zero-determinant strategies in repeated games \cite{press:2012:PNAS,hilbe:2013:PNAS,hilbe:2015:GEB}, we take a different approach by exploring outcome-contingent commitments in one-shot games.
In this work, we build upon the idea of extracting profit from players as presented by Bono and Wolpert \cite{bono:2014:EPE}. In contrast to their work, we do not treat the interaction as a bargaining problem; instead, we use classical game-theoretic analysis to develop a general non-coercive extortion scheme, where sequentiality is introduced into an otherwise simultaneous-move one-shot game.

We first derive the conditions under which this form of extortion is successful and, from these, identify the class of games that can be exploited using this scheme. Additionally, we determine both the maximum profit the extortioner can extract and the minimum payment they must offer for the extortion to succeed. We examine different role assignments within the extortion framework, where the extortioner can be either an external agent or a player within the game, and illustrate these scenarios using 2×2 games. Finally, we discuss potential applications of the proposed non-coercive extortion scheme that not only broadens our understanding of extortion mechanisms, it also raises new questions about how to safeguard strategic interactions from manipulation.

\vspace{-2mm}
\section{Formulation}
\label{sec:formulation}
We present a non-coercive extortion scheme in which the extortioner leverages the threat of an outcome-contingent payment to extract a profit from a game's player. This threat alters the game in a way that disadvantages the extorted player. As a result, the extorted player has an incentive to pay a fee to the extortioner to prevent an unfavorable modification to the equilibrium.

\subsection{Base game}

\begin{definition}[Base game]
The base game is an n-player finite normal-form game, $G_B = (N, S, u)$, where both the extorted player, $i_c$, and the recipient of the payment, $i_p$, are included in the set of players $N:=\{1,2,..,n\}$. A Nash equilibrium of this game is the strategy profile denoted as $s^*=(s_1^*, s_2^*, ..., s_n^*)$, where the set of Nash equilibria is $S^*$.
\end{definition}

\subsection{Binding threat}
\label{sec:binding_threat}

\begin{definition}[Binding threat]
    Let $G_B$ be the base game. The extortioner presents the following binding commitment to the extorted player, $i_c$:\\ 
    \textit{``Either you pay me a fee, or if you decline, I distort the game to your disadvantage".}\\
    More precisely, the threat has the following structure:\\
    \textit{``Either you pay me $c_1$, or if you decline, I give $c_2$ to player $i_p$ whenever they play strategy $s_{i_p}^{\circledast}$".}
    
where:
\begin{itemize}
    \item $c_1 > 0$, $c_1 \in \mathbb{R}$, is the profit the extortioner seeks to extract from player $i_c$.
    \item $c_2 > 0$, $c_2 \in \mathbb{R}$, is the outcome-contingent payment the extortioner commits to give to player $i_p$.
    \item The objective of the outcome-contingent payment is to shift $G_B$'s Nash equilibrium, therefore $s_{i_p}^\circledast \neq s_{i_p}^*$.
\end{itemize}
\label{def:binding_threat}
\end{definition}

\begin{remark}
    In an $n$-player game, the binding threat fully specifies the outcome-contingent payment when it depends on the decisions of $n-1$ players. Consequently, for a two-player base game ($G_B$), the binding threat described in definition \ref{def:binding_threat} is fully specified, as it only depends on the decisions of a single player, i.e. player $i_p$. For simplicity, this study focuses on the two-player framework. However, the analysis can be extended to an $n$-player game by allowing the binding threat to depend on additional players.
\end{remark}

The extortion scheme is considered non-coercive because it relies on a commitment to a reward ($c_2>0$), instead of a punishment. This commitment functions as a threat when it ultimately leads to a worse outcome for the extorted player. For the threat to be credible, it must be binding —meaning that if the extorted player declines, the extortioner still follows through with the promised payment, despite having no rational incentive to do so \cite{selten:1965:JITE}. Once this commitment is made, it becomes common knowledge.

\begin{figure}[h]
\centering
\includegraphics[width=0.49\textwidth]{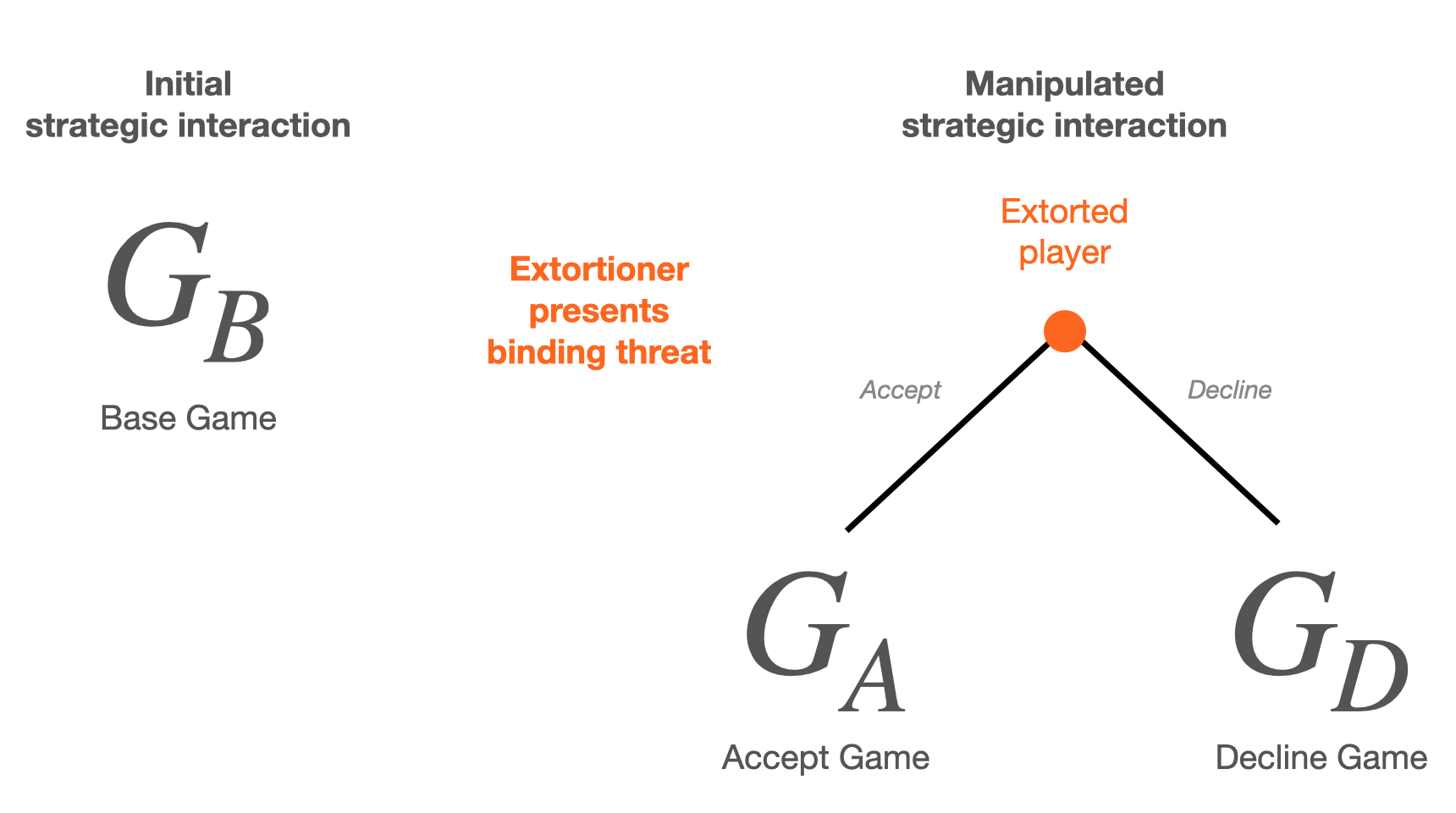}
\caption{\textbf{Extortioner's effect on the base game}. The strategic interaction between players is initially described by the base game ($G_B$). The extortioner presents a binding threat to the extorted player, who has to choose which game unfolds: the Accept game ($G_A$) or the Decline game ($G_D$). The resulting manipulated strategic interaction is an extensive-form game, denoted as $\Gamma_M$.}
\label{fig:general_scheme}
\end{figure}

By introducing this binding threat, the extortioner transforms the strategic interaction from a normal-form game ($G_B$) into an extensive-form game ($\Gamma_M$). In this new structure, the extorted player must first choose to Accept or Decline. This decision determines which game unfolds: Accept leads to $G_A$, while Decline leads to $G_D$, as shown in Figure \ref{fig:general_scheme}.

\begin{definition}(Manipulated game)
    Let $G_B$ be the normal-form game that characterises the initial strategic interaction among players, referred to as the base game. Suppose an agent, the extortioner, introduces a binding threat, thereby manipulating the structure of the interaction. This manipulation results in an extensive-form game, denoted by $\Gamma_M$, whose game tree is given in figure $\ref{fig:general_scheme}$, where
    \begin{enumerate}
        \item The extortioner presents the binding threat in reference to the base game ($G_B$)
        \item The extorted player, $i_c$, makes an initial decision between Accept and Decline
        \item If the extorted player chooses Accept, the game proceeds as $G_A$. Alternatively, if the extorted player chooses Decline, the game proceeds as $G_D$.
        \item Players engage in the chosen game, $G_A$ or $G_D$.
    \end{enumerate}

    Thus, the strategic interaction ultimately unfolds within $\Gamma_M$, which results from the manipulation imposed on the base game by the extortioner.
\end{definition}

\subsection{Accept game}
Once the extortioner presents the binding threat, the Accept game ($G_A$) unfolds if the extorted player, $i_c$, accepts to pay $c_1$ to the extortioner, resulting in a successful extortion.

\begin{definition}[Accept game]
    The Accept game is a modification of the base game, where $c_1$ is subtracted from the extorted player's payoffs for all strategy profiles, i.e.
    \begin{equation}
        G_A = (N,S,u^{A})
        \label{eq:gen_acceptgame}
    \end{equation}
    where: 
    \begin{equation}
    \begin{aligned}
    u_{i_c}^{A}(s)= u_{i_c}(s) - c_1, \hspace{5mm} \forall s \in S \\
    u_{-{i_c}}^{A}(s)= u_{-{i_c}}(s), \hspace{5mm} \forall s \in S.
    \end{aligned}
    \label{eq:trans_acceptgame}
\end{equation} 
\label{def:accept_game}
\end{definition}

A Nash equilibrium of the Accept game is a strategy profile denoted as $s^{*A}=(s_1^{*A}, s_2^{*A}, ..., s_n^{*A})$, where the set of Nash equilibria is $S^{*A}$.

\vspace{5mm}

It can be noted that the Accept game transformation ($u^{A}$) is an affine transformation. Thus, it preserves the underlying structure of the game, ensuring that the relative ranking of payoffs remains unchanged with respect to the base game. As a result, the Nash equilibria set of the base game ($S^*$) is equivalent to the Accept game set ($S^{*A}$).

\subsection{Decline game}
Once the extortioner presents the binding threat, the Decline game ($G_D$) unfolds if the extorted player, $i_c$, declines to pay $c_1$ to the extortioner, resulting in the extortioner effecting the outcome-contingent payment to player $i_p$.

\begin{definition}[Decline game]
    The Decline game is a modification of the base game, where $c_2$ is added to player $i_p$'s payoffs only for strategy $s_{i_p}^{\circledast}$, i.e.
    \begin{equation}
        G_D = (N,S,u^{D})
        \label{eq:gen_acceptgame}
    \end{equation}
    where: 
    \begin{equation}
    \begin{aligned}
    u_{i_p}^{D}(s)= u_{i_p}(s) + c_2 \mathds{1}_{\{s_{i_p} = s_{i_p}^{\circledast} \}}, \hspace{5mm} \forall s \in S \\
    u_{-{i_p}}^{D}(s)= u_{-{i_p}}(s), \hspace{5mm} \forall s \in S.
    \end{aligned}
    \label{eq:trans_declinegame}
\end{equation} 
\label{def:decline_game}
\end{definition}

A Nash equilibrium of the Decline game is a strategy profile denoted as $s^{*D}=(s_1^{*D}, s_2^{*D}, ..., s_n^{*D})$, where the set of Nash equilibria $S^{*D}$.\\

By construction, we define $\Gamma_M$ as an extensive-form game with perfect recall and complete information. However, since $G_A$ and $G_D$ are normal-form games where decisions are made simultaneously, not all information sets are singletons. Thus, $\Gamma_M$ is an extensive-form game with complete but imperfect information.

\section{Conditions for a successful extortion}
\label{sec:conditions_successful}

Extortion is successful when the extortioner extracts a profit from the extorted player, $i_c$, namely when $i_c$ accepts to pay $c_1$ to the extortioner. We derive the key conditions for this to occur. From these conditions, we identify the class of games that are susceptible to this scheme. Additionally, we determine the maximum profit the extortioner can extract ($c_1^{max}$) and the minimum payment required for the extortion to succeed ($c_2^{min}$).

The first condition is ensuring a unique pure Nash equilibrium in the base game ($G_B$), as well as in the Accept game ($G_A$), and Decline game ($G_D$). This prevents an equilibrium selection problem, where the existence of multiple equilibria leads to strategic uncertainty in players' decision-making \cite{harsanyi:1988:book,vanDamme:1991:book,moore:2025:workingpaper}. Similarly, we focus on pure Nash equilibria to further eliminate the potential for such uncertainty, as discussed in section \ref{sec:discussion}.

\begin{assumption}
    The base game ($G_B$) has a unique pure Nash equilibrium, i.e. $S^*$ is a singleton.
    \label{assump:uniqueNash_base}
\end{assumption}

Given this assumption, it can also be established that $G_A$ has a unique pure Nash equilibrium. Next, we derive a theorem consisting of two conditions on $c_2$ to ensure that $G_D$ has a unique pure Nash equilibrium resulting from the extortioner's outcome-contingent payment. The first condition requires $c_2$ to be large enough to create a new Nash equilibrium ($s^{*D}$) in which player $i_p$ selects $s_{i_p}^{\circledast}$. The second condition ensures that $c_2$ is sufficiently large for $s^{*D}$ to be the unique equilibrium of $G_D$, eliminating any ambiguity in player $i_p$'s decisions.

\begin{theorem}
    Let $G_B$ be a two-player game. The Decline game $G_D$ arises from the commitment of the extortioner to pay $c_2$ to player $i_p$ whenever they play strategy $s_{i_p}^{\circledast}$.
    The Decline game has a unique pure Nash equilibrium ($s^{*D}$) satisfying $s_{i_p}^{\circledast} \in s^{*D}$, where $s_{i_p}^{\circledast} \neq s_{i_p}^* $, if and only if the following two conditions hold:
\begin{enumerate}
    \item Equilibrium creation condition:  $$\hspace{-2mm} c_2 > u_{i_p}(s_{i_p}, s_{-i_{p}}^{*D}) - u_{i_p}(s^{*D}), \hspace{1mm} \forall s_{i_{p}} \in S_{i_{p}} \hspace{1mm} \textnormal{where} \hspace{1mm} s_{i_p} \neq s_{i_p}^{\circledast}$$
    \item Equilibrium uniqueness condition: $$ c_2 > u_{i_p}(s^*) - u_{i_p}(s_{i_p}^{\circledast}, s_{-i_p}^*)$$
\end{enumerate}
\label{def:theorem_2}
\end{theorem}

\begin{proof}
    Let $G_B$ a two-player game. The best response for player $-i_p$ to $s_{i_p}^{\circledast}$ is

    \begin{equation*}
    \begin{split}
        B_{-i_p}(s_{i_p}^{\circledast}) = \{s_{-i_p}^{*D} \in S_{-i_p} | u_{-i_p}(s_{-i_p}^{*D},s_{i_p}^{\circledast}) > u_{-i_p}(s_{-i_p}, s_{i_p}^{\circledast}), \\
        \forall s_{-i_p} \in S_{-i_p}\ \hspace{1mm} \textnormal{where} \hspace{1mm} s_{-i_p} \neq s_{-i_p}^{*D} \}
    \end{split}
    \end{equation*}

    The inequality is strict to avoid the emergence of multiple equilibria. Once $s_{-i_p}^{*D}$ is established, the condition $B_{i_p}(s_{-i_p}^{*D})= s_{i_p}^{\circledast}$ is imposed, as follows

    \begin{equation}
    \begin{split}
        B_{i_p}(s_{-i_p}^{*D}) = \{s_{i_p}^{\circledast} \in S_{i_p} | u_{i_p}^D(s_{i_p}^{\circledast}, s_{-i_p}^{*D}) > u_{i_p}^D(s_{i_p},  s_{-i_p}^{*D}), \\ \forall s_{i_p} \in S_{i_p} \hspace{2mm} \textnormal{where} \hspace{1mm} s_{i_p} \neq s_{i_p}^{\circledast}\}.
    \end{split}
        \label{eq:theorem_2}
    \end{equation}

    Once again, the inequality is strict such that there is a single best response. \\
    \textit{Equilibrium creation condition}: as every player choose their best response in a Nash equilibrium, there exists a Nash equilibrium for the Decline game $G_D$ satisfying $s_{i_p}^{\circledast} \in s^{*D}$, where $s^{*D} =(s_{i_p}^{\circledast}, s_{-i_{p}}^{*D})$. From equations \ref{eq:trans_declinegame} and \ref{eq:theorem_2}, the condition on $c_2$ can be written as
    $$u_{i_p}(s^{*D}) + c_2 > u_{i_p}(s_{i_p},  s_{-i_{p}}^{*D}), \hspace{3mm} \forall s_{i_p} \in S_{i_p} \hspace{2mm} \textnormal{where} \hspace{1mm} s_{i_p} \neq s_{i_p}^{\circledast}.$$

    \textit{Equilibrium uniqueness condition}: to ensure that $s^{*D}$ is unique, then $u_{i_p}^D(s_{i_p}^{\circledast}, s_{-i_p}^*) > u_{i_p}(s^{*})$. This condition guarantees that $s^*$ is not a Nash equilibrium of the Decline game $G_D$.
\end{proof}

After establishing equilibrium uniqueness for $G_B$, $G_A$ and $G_D$, we derive a theorem that guarantees the success of the extortion. Since the extortion scheme relies on a binding threat that shifts the equilibrium to a worse outcome for the extorted player, we must ensure that once the threat is introduced, the Nash equilibrium of the Decline game ($s^{*D}$) leads to a worse outcome for the extorted player, $i_c$, than the equilibrium of the Accept game ($s^{*A}$). This makes selecting the Accept game the rational choice for $i_c$, leading them to pay $c_1$ to the extortioner, ensuring the success of the scheme.

\begin{theorem}
    Let $\Gamma_M$ be the manipulated game. Suppose that the Nash equilibria sets of $G_A$ and $G_D$ are singletons. Then, the extortioner obtains a profit if and only if the following condition holds for the extorted player
    \begin{equation}
        u_{i_c}^{A}(s^{*A}) > u_{i_c}^{D}(s^{*D}).
        \label{eq:condition_extortion}
    \end{equation}
    \label{def:theorem_1}
\end{theorem}

\begin{proof}
    The extortioner profits only if player $i_c$ chooses Accept at the root of $\Gamma_M$'s game tree. To determine the player's decision, we apply the subgame perfect equilibrium (SPE) criterion, the standard solution concept for extensive games, which ensures a Nash equilibrium in every subgame of $\Gamma_M$. Since the subgames of $\Gamma_M$ are $G_A$, $G_D$ and $\Gamma_M$ itself, player $i_c$ will choose Accept if the Nash equilibrium of the Accept game ($s^{^*A}$) yields a strictly higher payoff than that of the Decline game ($s^{^*D}$).
\end{proof}

\subsection{Susceptible games}

Based on the conditions for successful extortion, we can identify the structure of base games ($G_B$) that are susceptible to extortion. Assumption \ref{assump:uniqueNash_base} narrows our focus to games with a unique pure Nash equilibrium, thereby excluding games with multiple equilibria —such as (anti-) coordination games— and those without any pure equilibrium, like cyclic games.

Moreover, for a game to be susceptible to extortion, an outcome-contingent reward must shift the Nash equilibrium to a worse payoff for the extorted player compared to the equilibrium of $G_B$. Only in such cases the extortioner can extract a profit by altering the equilibrium. This requirement leads directly to the following condition, derived from theorem \ref{def:theorem_1} and definition \ref{def:binding_threat}.

\begin{corollary}
    Let $G_B$ be the base game. The extortioner can extract a profit from the base game if and only if the following condition holds for the extorted player
    \begin{equation}
        u_{i_c}(s^*)>u_{i_c}(s^{*D})        
    \end{equation}
    where it is assumed that $s_{i_p}^{\circledast} \in s^{*D}$.
\label{def:susceptible_games}
\end{corollary}

\subsection{Maximum extractable profit}

\begin{corollary}
    Let $G_B$ be a susceptible game. If the extorted player, $i_c$, decides to Accept, then the maximum profit that can be extracted by the extortioner is 
    \begin{equation}
    \begin{split}
        c_1^{max}=\sup\{c_1 >0 | u_{i_c}(s^{*A}) - u_{i_c}^{D}(s^{*D}) > c_1\}, \\ \text{where} \quad c_1^{max} > c_1.
        \label{eq:condition_maxgain}
    \end{split}
    \end{equation}
    \label{def:max_profit}
\end{corollary}

\begin{proof}
    From definition \ref{def:accept_game} and theorem \ref{def:theorem_1}, it follows that
    \begin{equation}
        u_{i_c}(s^{*A}) - c_1 > u_{i_c}^{D}(s^{*D}).
        \label{eq:condition_maxgain}
    \end{equation}
    This condition establishes the maximum value where $c_1^{max} > c_1 \in \mathbb{R}$. If $c_1^{max} = c_1$, then it is not guaranteed that the extorted player Accepts the offer.
\end{proof}

\subsection{Minimum required payment}
\begin{corollary}
    Let $G_B$ be a susceptible game. The minimum required payment, $c_2^{min}$, that the extortioner must offer player $i_p$ to successfully shift the Nash equilibrium of $G_B$ is given by the supremum of the set of $c_2$ values satisfying the conditions of theorem \ref{def:theorem_2}, where $c_2 > c_2^{min}$.
    \label{def:min_pay}
\end{corollary}

\section{Extortion scenarios and examples}

\begin{figure}[h!]
\centering
\includegraphics[width=0.49\textwidth]{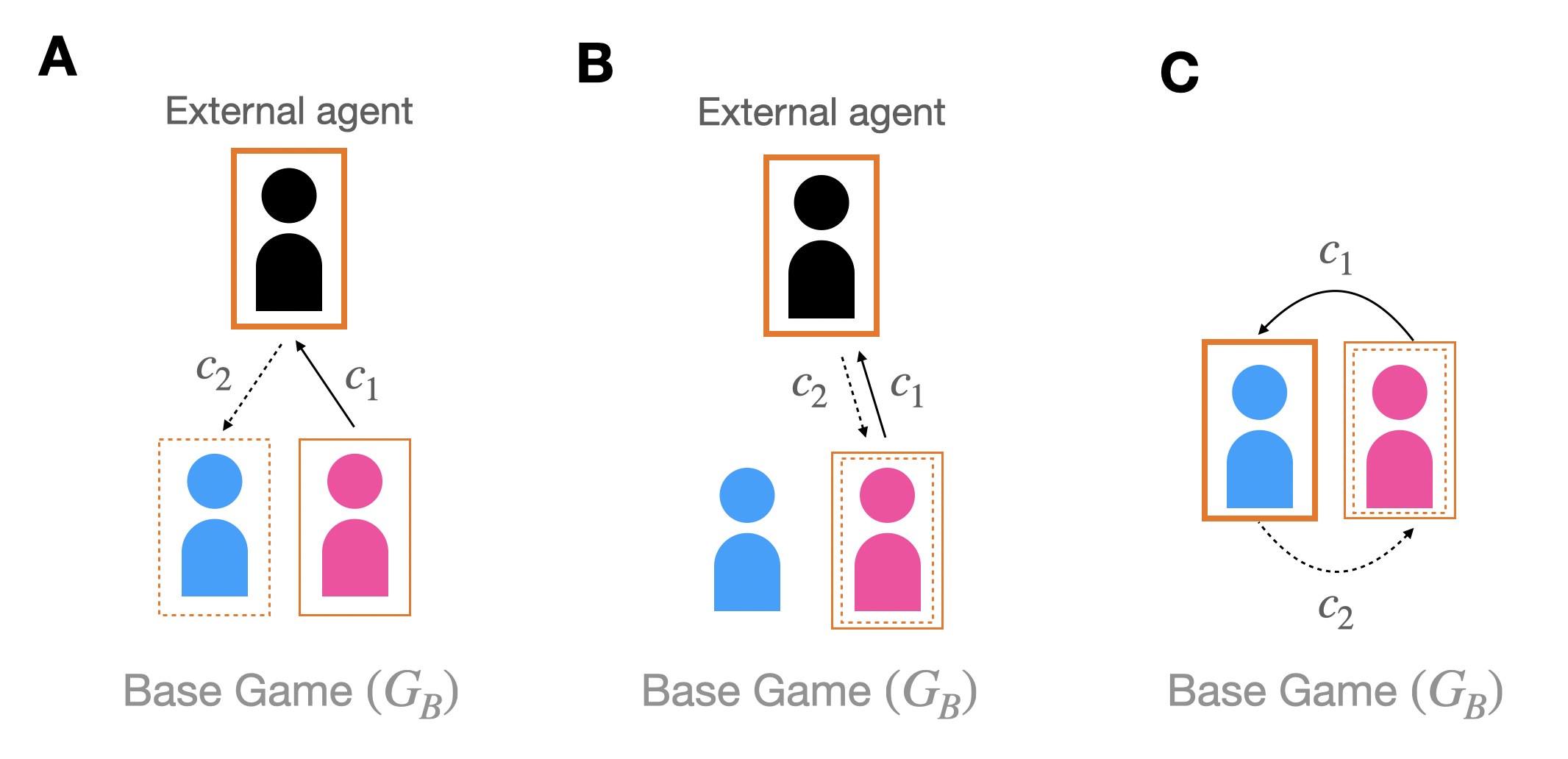}
\caption{\textbf{Extortion scenarios.} The extortion scheme involves three distinct roles: the extortioner, the extorted player ($i_c$) and the recipient of the payment ($i_p$). The extortioner aims to extract a profit ($c_1$) from $i_c$ by posing the threat of making an outcome-contingent payment ($c_2$) to $i_p$. The specific extortion scenario depends on how these roles are assigned and whether the extortioner is an external agent or a player within the game. The bold-bordered rectangle indicates the extortioner, the thin-bordered rectangle surrounds the extorted player $i_c$, and the dashed-bordered rectangle indicates the recipient of the payment $i_p$. \textbf{A.} Two-player extortion with external agent. \textbf{B.} One-player extortion with external agent. \textbf{C.} One-player extortion without external agent.}
\label{fig:2_player_scheme}
\end{figure}

The extortion scheme involves three distinct roles: the extortioner, the extorted  player ($i_c$) and the recipient of the payment ($i_p$). The possible extortion scenarios depend on how these roles are assigned and whether the extortioner is an external agent or a player within the game. Figure \ref{fig:2_player_scheme} illustrates the possible non-coercive extortion options for a two-player base game. We first examine the case where the extortioner is an external agent. If $i_c$ and $i_p$ are assigned to different players, the scheme constitutes a two-player extortion. Alternatively, if the same player is both extorted and receives the payment, the scheme operates as a one-player extortion. Finally, we consider the case where the extortioner is a player within the game.

\subsection{Two-player extortion with external agent}
\label{sec:twoplayer_extortion}
We analyse a two-player base game composed of players $i$ and $j$. We begin with the scenario illustrated in figure \ref{fig:2_player_scheme}A, where the extorted player and the payment recipient are different players ($i_c \ne i_p$), and an external agent acts as the extortioner. In this setup, the extortioner presents the following binding threat to player $i$:
\textit{``Either give me $c_1$, or if you decline, I give $c_2$ to your co-player whenever they play strategy $s_j^{\circledast}$".}

Here, player $i$ is the extorted player ($i=i_c$) and player $j$ is the recipient of the outcome-contingent payment ($j=i_p$).
Theorems \ref{def:theorem_2} and \ref{def:theorem_1} indicate the conditions for a successful extortion. In particular, from corollary \ref{def:max_profit} we can derive a more direct expression for the maximum extractable profit.

\begin{proposition}[Two-player extortion - Maximum extractable profit]
    Let $G_B$ be a susceptible game. In a two-player extortion, the maximum extractable profit is
    \begin{equation}
    \begin{split}
        c_1^{max} = \sup\{c_1 >0| u_{i_c}(s^{*A})-u_{i_c}(s^{*D})>c_1 \}, \\ \text{where} \quad c_1^{max} > c_1.
    \end{split}
    \end{equation}
    \label{def:two-player_maxprofit}
\end{proposition}

\subsubsection{2x2 Games}

\begin{figure}[h!]
\centering
\includegraphics[width=14cm]{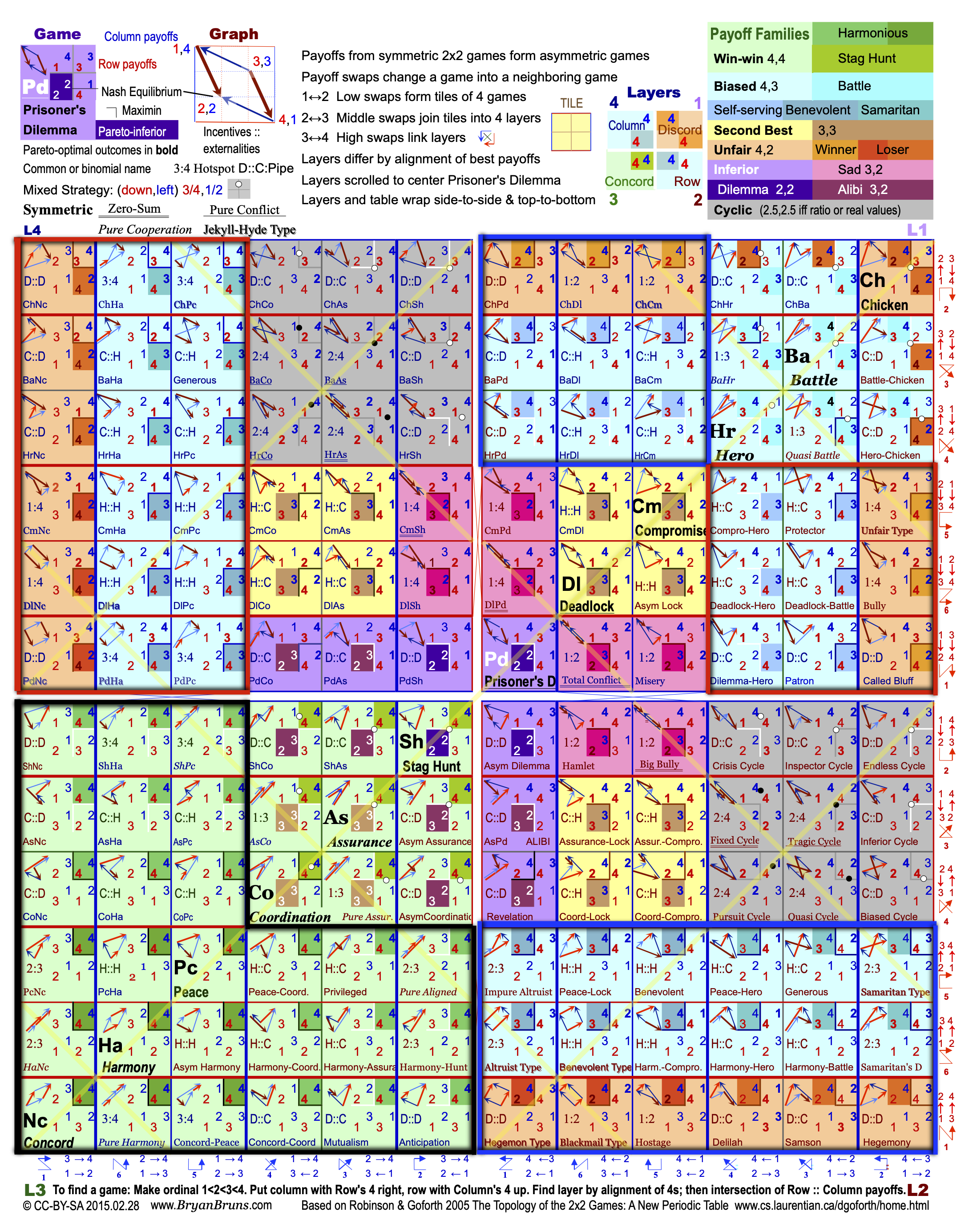}
\caption{\textbf{2x2 games susceptible to a two-player extortion}. To indicate the class of games that can be exploited, we use the periodic chart of 2x2 games \cite{robinson:2005:book,bruns:2015:atlas}. According to corollary \ref{def:susceptible_games}, the games outlined with a bold red border indicate cases where the row player is susceptible to extortion, while those with a bold blue border indicate susceptibility for the column player. Games enclosed by a bold black border represent cases where both players are susceptible.
To simplify the analysis and in accordance with assumption \ref{assump:uniqueNash_base}, we focus on games with a unique pure strategy Nash equilibrium, excluding (anti-)coordination and cyclic games. Under these constraints, $56,25\%$ of all 2x2 games are susceptible to the two-player extortion scheme. These are the games where at least one player achieves their best possible outcome in equilibrium, making them the vulnerable party. As a result, win-win, biased, and unfair games are susceptible to extortion, with win-win games being the most vulnerable,  since either player can be targeted by the extortioner.}
\label{fig:twoplayer_susceptible}
\end{figure}

To illustrate the extortion scheme, we focus on the simplest strategic interactions: two-player, two-action (2x2) games. To categorise these games systematically, we refer to the periodic chart of 2x2 games \cite{robinson:2005:book,bruns:2015:atlas}, which classifies games based on their ordinal payoff patterns. Thus, if payoffs are randomly assigned, different types of ordinal games occur in the proportions reflected in this chart (see Figs. \ref{fig:twoplayer_susceptible} and \ref{fig:oneplayer_susceptible}). In particular, we focus on base games with a unique pure Nash equilibrium, in line with assumption \ref{assump:uniqueNash_base}. This excludes games where both players lack a dominant strategy, eliminating $25\%$ of all 2x2 games. We find that a game is susceptible to the two-player extortion scheme if at least one player achieves their best possible outcome in equilibrium, making them the vulnerable party to extortion. To determine such games, we analyse how an outcome-contingent payment to the co-player shifts the Nash equilibrium.

\begin{itemize}
    \item Case 1: The extorted player ($i_c$) has a dominant strategy.
\end{itemize}
If $i_c$ has a dominant strategy, their strategy remains unchanged when the payment to the co-player is introduced. Only the co-player's strategy shifts, altering the Nash equilibrium along a single axis: $s^*=(s_i^*,s_j^*)$ to $s^{*D}=(s_i^{*D},s_j^{\circledast})$, where $s_j^* \neq s_j^{\circledast}$ and $s_i^* = s_i^{*D}$. For instance, if the row player is extorted, the payment shifts the equilibrium column-wise but not row-wise.

\begin{itemize}
    \item Case 2: The extorted player ($i_c$) does not have a dominant strategy.
\end{itemize}
Here, the strategies of both the extorted player and co-player change, causing the equilibrium to shift diagonally: $s^*=(s_i^*,s_j^*)$ to $s^{*D}=(s_i^{*D},s_j^{\circledast})$, where $s_j^* \neq s_j^{\circledast}$ and $s_i^* \neq s_i^{*D}$. For example, if the row player is extorted, the payment shifts the equilibrium both column-wise and row-wise. Note that in this case the co-player must have a dominant strategy for the base game to have a unique pure Nash equilibrium.

By analysing these shifts, we can systematically determine which games allow for a successful extortion as shown in figure \ref{fig:twoplayer_susceptible}. We find that $56,25\%$ of all 2x2 games are susceptible to the two-player extortion scheme. These are the games where at least one player achieves their best possible outcome in equilibrium, making them the vulnerable party. As a result, win-win, biased, and unfair games are susceptible, with win-win games being the most vulnerable, since either player can be targeted for extortion.

\vspace{-5mm}
\subsubsection{Example}

A natural example to illustrate the two-player extortion scheme is a win-win (no-conflict) game, where both players achieve their best possible outcome in equilibrium. One such game is \textit{Concord}, a symmetric game in which both players have dominant strategies that lead to mutual benefit \cite{bruns:2015:atlas}. In this structure, cooperation aligns with individual incentives, making it the antithesis of the Prisoner's Dilemma. Building on this, we illustrate the two-player extortion scheme using a base game with a Concord structure.

\begin{table}[!h]
    \centering
    \caption*{Base game ($G_B$)}
    \begin{tabular}{c|c|c|}
    &L&R \\
    \hline
    T& 4,4&8,6\\
    \hline
    B& 6,8& \textcolor{violet}{10,10} \\
    \hline
    \end{tabular}
    \label{tab:basegame}
\end{table}

The external agent presents the following binding threat to the row player: \textit{``Either give me 2 (Euros), or if you decline, I give 3 to the column player whenever they play Left."}
\begin{table}[!h]
\caption*{Manipulated game ($\Gamma_M$)}
\begin{subtable}[c]{0.23\textwidth}
\centering
\begin{tabular}{c|c|c|}
    &L&R \\
    \hline
    T& 2,4&6,6\\
    \hline
    B& 4,8& \textcolor{violet}{8,10} \\
    \hline
\end{tabular}
\subcaption*{Accept game ($G_A$)}
\end{subtable}
\begin{subtable}[c]{0.23\textwidth}
\centering
\begin{tabular}{c|c|c|}
    &L&R \\
    \hline
    T& 4,7&8,6\\
    \hline
    B& \textcolor{violet}{6,11}& 10,10 \\
    \hline
\end{tabular}
\subcaption*{Decline game ($G_D$)}
\end{subtable}
\end{table}

The binding threat transforms $G_B$ into an extensive game ($\Gamma_M$), where the row player must decide between playing the Accept or the Decline game. Given that the Nash equilibrium of $G_A$ provides a better payoff for the row player compared to $G_D$, the rational decision for the row player is to Accept paying the extortion to the external agent.
According to corollary \ref{def:max_profit} and the related proposition \ref{def:two-player_maxprofit}, the maximum extractable profit is $c_1^{max}=4$, with $c_1^{max}>c_1$. Similarly, corollary \ref{def:min_pay} states that the minimum required payment is $c_2^{min}=2$ , where $c_2 > c_2^{min}$.

\subsection{One-player extortion with external agent}
\label{sec:oneplayer_external}
Figure \ref{fig:2_player_scheme}B illustrates the scenario where the extorted player and the payment recipient are the same player ($i_c = i_p$), and an external agent acts as the extortioner. In this setup, the extortioner presents the following binding threat to player $i$:\\
\textit{``Either give me $c_1$, or if you decline, I give you $c_2$ whenever you play strategy $s_i^{\circledast}$".}

Here, player $i$ assumes both roles as the extorted player and the recipient of the outcome-contingent payment ($i=i_c=i_p$). This scenario highlights the counterintuitive effect of how receiving a reward can ultimately lead a player to a worse equilibrium. Notably, this effect makes the one-player scheme more restrictive compared to its two-player counterpart, as shown in the following proposition.

\begin{proposition}[One-player extortion restriction I]
    Let $G_B$ be a two-player game. An external agent presents a one-player extortion scheme where player $i$ is both the extorted player and the payment recipient. The payment aims to shift the Nash equilibrium from $s^*=(s_i^*,s_j^*)$ to $s^{*D}=(s_i^{\circledast}, s_j^{*D})$, where $s_i^* \neq s_i^{\circledast}$. The one-player extortion scheme is feasible only if $s_{j}^{*D} \neq s_{j}^*$.
    \label{def:one-player_restriction_I}
\end{proposition}

\begin{proof}
    By theorem \ref{def:theorem_1}, we have the inequality: $$u_i(s^{*A}) - u_i(s^{*D}) -c_1>c_2.$$ Additionally, theorem \ref{def:theorem_2} imposes the following condition on $c_2$: $$c_2 > u_i({s^*})-u_i(s_i^{\circledast},s_j^*).$$
    Given that $s^{*A}=s^*$, then the combination of the two inequalities can be written as:
    \begin{equation}
        u_i(s^{*}) - u_i(s^{*D}) -c_1>c_2>u_i({s^*})-u_i(s_i^{\circledast},s_j^*).
    \end{equation}
    Since $s^{*D}=(s_i^{\circledast},s_j^{*D})$, the inequality becomes inconsistent if $s_{j}^{*D}=s_{j}^*$. Therefore, for the one-player extortion scheme to be feasible,  $s_{j}^{*D} \neq s_{j}^*$ must hold, restricting the class of games susceptible to extortion.
\end{proof}

\begin{corollary}
    Let $G_B$ be a two-player game where $s^*=(s_i^*,s_j^*)$ is the unique pure Nash equilibrium. A one-player extortion scheme is feasible if and only if player $j$ does not have a dominant strategy.
    \label{def:one-player_corollary}
\end{corollary}

\begin{proof}
    If $s_j^*$ is a dominant strategy for player $j$, it is the best response for all of player $i$'s strategies. Consequently, if the outcome-contingent payment shifts player $i$'s strategy from $s_i^*$ to $s_i^\circledast$, player $j$'s best response would remain the same in both cases, leading to $s_j^{*D} = s_j^*$. This contradicts proposition \ref{def:one-player_restriction_I}, making the extortion infeasible.
\end{proof}

Using corollary \ref{def:max_profit}, we derive a more direct expression for the maximum extractable profit. Comparing this result with proposition \ref{def:two-player_maxprofit} shows that the two-player extortion is more profitable than the one-player extortion scheme.

\begin{proposition}[One-player extortion - Maximum extractable profit]
    Let $G_B$ be the base game. In a one-player extortion, the maximum extractable profit is
    \begin{equation}
    \begin{split}
        c_1^{max} = \sup\{c_1 >0| u_{i_c}(s^{*A})-u_{i_c}(s^{*D})-c_2>c_1 \}, \\ \text{where} \quad c_1^{max} > c_1,
    \end{split}
    \end{equation}
    for a valid value of $c_2$.
    \label{def:one-player_maxprofit}
\end{proposition}

Notably, in the one-player extortion, the allowed ranges for $c_1$ and $c_2$ become interdependent, further constraining the scheme.

\begin{proposition}[One-player extortion restriction II]
    Let $G_B$ be the base game. The extortioner can extract a profit from the base game if and only if the following condition holds for the extorted player
    \begin{equation}
        u_{i_c}(s^{*A})-u_{i_c}(s^{*D})>c_2^{min}.
    \end{equation}
    \label{def:one-player_restriction_II}
\end{proposition}

\begin{proposition}[One-player extortion - Maximum allowed payment]
    Let $G_B$ be the base game. In a one-player extortion, the maximum allowed payment for a successful extortion is
    \begin{equation}
    \begin{split}
        c_2^{max} = \sup\{c_2 >0| u_{i_c}(s^{*A})-u_{i_c}(s^{*D})-c_1>c_2 \}, \\ \text{where} \quad c_2^{max} > c_2,
    \end{split}
    \end{equation}
    for a valid value of $c_1$.
    \label{def:one-player_maxpayment}
\end{proposition}

\subsubsection{2x2 Games}

The set of 2x2 games susceptible to a one-player extortion is a subset of those vulnerable to the two-player extortion scheme. In this more restrictive subset, the extorted player must have a dominant strategy, while their co-player must not. Additionally, the extorted player’s best possible outcome must lie diagonally opposite their worst outcome in the payoff matrix.

This added restrictiveness stems from the structure of the one-player extortion scheme, where the extorted player is also the recipient of the payment. As a result, the conditions required to shift the Nash equilibrium are more stringent. 

For theorem \ref{def:theorem_1} to hold, the outcome-contingent payment must induce a shift from $s^*=(s_i^*,s_j^*)$ to $s^{*D}=(s_i^{\circledast}, s_j^{*D})$, where $s_i^* \neq s_i^{\circledast}$. According to proposition \ref{def:one-player_restriction_I}, this shift also requires $s_j^{*D} \neq  s_j^*$, meaning that the co-player cannot have a dominant strategy, as shown in corollary \ref{def:one-player_corollary}. As a result, the extorted player must have a dominant strategy to ensure that the game has a unique pure Nash equilibrium. This restricts feasible equilibrium shifts to diagonal moves only.
Moreover, proposition \ref{def:one-player_restriction_II} imposes a further constraint: only games in which the extorted player’s best outcome lies diagonally from their worst remain susceptible to extortion. These restrictions are visually illustrated in figure \ref{fig:oneplayer_susceptible}. Under these conditions, we find that $16,\overline{6}\%$ of all 2x2 games are susceptible to the one-player extortion scheme.

\subsubsection{Example}

To illustrate this scenario, we consider an asymmetric Concord base game that meets the susceptibility criteria for a one-player extortion


\begin{table}[!h]
    \centering
    \caption*{Base game ($G_B$)}
    \begin{tabular}{c|c|c|}
    &L&R \\
    \hline
    T& 4,6&8,4\\
    \hline
    B& 6,8& \textcolor{violet}{10,10} \\
    \hline
    \end{tabular}
    \label{tab:basegame}
\end{table}

The external agent presents the following binding threat to the row player: \textit{``Either give me 2 (Euros), or if you decline, I give you 3 whenever you play Top."}

\begin{table}[!h]
\caption*{Manipulated game ($\Gamma_M$)}
\begin{subtable}[c]{0.23\textwidth}
\centering
\begin{tabular}{c|c|c|}
    &L&R \\
    \hline
    T& 2,6&6,4\\
    \hline
    B& 4,8& \textcolor{violet}{8,10} \\
    \hline
\end{tabular}
\subcaption*{Accept game ($G_A$)}
\end{subtable}
\begin{subtable}[c]{0.23\textwidth}
\centering
\begin{tabular}{c|c|c|}
    &L&R \\
    \hline
    T& \textcolor{violet}{7,6}&11,4\\
    \hline
    B& 6,8& 10,10 \\
    \hline
\end{tabular}
\subcaption*{Decline game ($G_D$)}
\end{subtable}
\end{table}

Given that the Nash equilibrium of $G_A$ provides a better payoff for the row player compared to $G_D$, the rational decision for the row player is to Accept paying the extortion to the external agent. According to corollary \ref{def:max_profit} and the related proposition \ref{def:one-player_maxprofit}, the maximum extractable profit is $c_1^{max}=3$, with $c_1^{max} > c_1$. Similarly, corollary \ref{def:min_pay} and proposition \ref{def:one-player_maxpayment} determine the allowed range for the payment: $c_2^{max}=4$ and $c_2^{min}=2$, where $ c_2^{max} > c_2 > c_2^{min}$.

\subsection{One-player extortion without external agent}

Finally, we examine the scenario where the extortioner is a player within the game, as shown in figure \ref{fig:2_player_scheme}C. In this one-player extortion, the extorted player and the payment recipient are the same player ($i_c=i_p=i$), while the extortioner is the co-player ($i_e=j$). Player $j$ presents the following binding threat to player $i$:\\
\textit{``Either give me $c_1$, or if you decline, I give you $c_2$ whenever you play strategy $s_i^{\circledast}$".}

Here, the binding threat acts as a commitment to a side-payment, where the extortioner's payoffs are reflected in the base game $G_B$.

\begin{definition}[Extortioner player's payoffs]
Let the extortioner be a player within base game $G_B$. The extortioner player, $i_e$, presents a binding threat to their co-player that alters $i_e$'s payoffs as follows: 
\begin{itemize}
    \item For the Accept game $G_A$, the extortioner player's payoff function is modified as
    \begin{equation}
        u_{i_e}^{A}(s)=u_{i_e}(s)+c_1, \hspace{5mm} \forall s \in S.
    \end{equation}
    \item For the Decline game $G_D$, the extortioner player's payoff function is modified as
    \begin{equation}
        u_{i_e}^{D}(s)=u_{i_e}(s)-c_2\mathds{1}_{\{s_{i_p} = s_{i_p}^{\circledast} \}}, \hspace{5mm} \forall s \in S.
    \end{equation}
\end{itemize}
\end{definition}

\begin{remark}
    The Nash equilibria $s^{*A}$ and $s^{*D}$ are invariant to the extortioner player's payoff modifications, as these modifications do not alter player $i_e$'s best responses.
\end{remark}

For the extortion to be successful, conditions in section \ref{sec:conditions_successful} and \ref{sec:oneplayer_external} must be fulfilled. In addition, for the extortion to be well-motivated, it must be worthwhile for the extortioner player, namely the Accept game yields a higher payoff than the Decline game.

\begin{proposition}[Extortioner player's condition]
    Let $G_B$ be the base game. If the extortioner is a player within $G_B$, a one-player extortion is worthwhile if and only if the following condition holds
    \begin{equation}
        u_{i_e}^{A}(s^{*A})>u_{i_e}^D(s^{*D}).
    \end{equation}
    \label{def:worthwhile_condition}
\end{proposition}

Moreover, the extortioner player's condition and theorem \ref{def:theorem_1} impose a constraint on the payoff magnitudes of the base game, narrowing the possible games where it is worthwhile for the extortioner player to present a one-player extortion to a co-player.

\begin{proposition}
    Let $G_B$ be the base game. It is worthwhile for the extortioner player, $i_e$, to present one-player extortion to a co-player, $i_c$, if and only if the following condition holds for $G_B$
    \begin{equation}
        u_{i_c}(s^{*A}) - u_{i_c}(s^{*D}) > u_{i_e}(s^{*D}) - u_{i_e}(s^{*A}).
    \end{equation}
    \label{def:worthwhile_games}
\end{proposition}

Proposition \ref{def:worthwhile_condition} also establishes a minimum profit that the extortioner player must extract for the one-player extortion to be worthwhile.

\begin{proposition}[Extortioner player's minimum profit]
    Let $G_B$ be the base game. In a one-player extortion where the extortioner is a player within a game, the minimum profit the extortioner player, $i_e$, must extract for the extortion to be worthwhile is
    \begin{equation}
    \begin{split}
        c_1^{min}=\inf\{c_1>0 | c_1 > u_{i_e}(s^{*D})-u_{i_e}(s^*)-c_2\}, \\ \text{where} \quad c_1 > c_1^{min}>0,
    \end{split}
    \label{eq:min_profit}
    \end{equation}
    for a valid value of $c_2$.
\label{def:min_profit}
\end{proposition}

As $c_1$ and $c_2$ are interdependent, a condition for the minimum payment ($c_2^{min}$) also arises from this proposition. In this scenario, $c_2^{min}$ is then determined as the maximum value satisfying both this condition and the requirements outlined in theorem \ref{def:theorem_2}.

\subsubsection{2x2 Games}

\begin{figure}[h!]
\centering
\includegraphics[width=14cm]{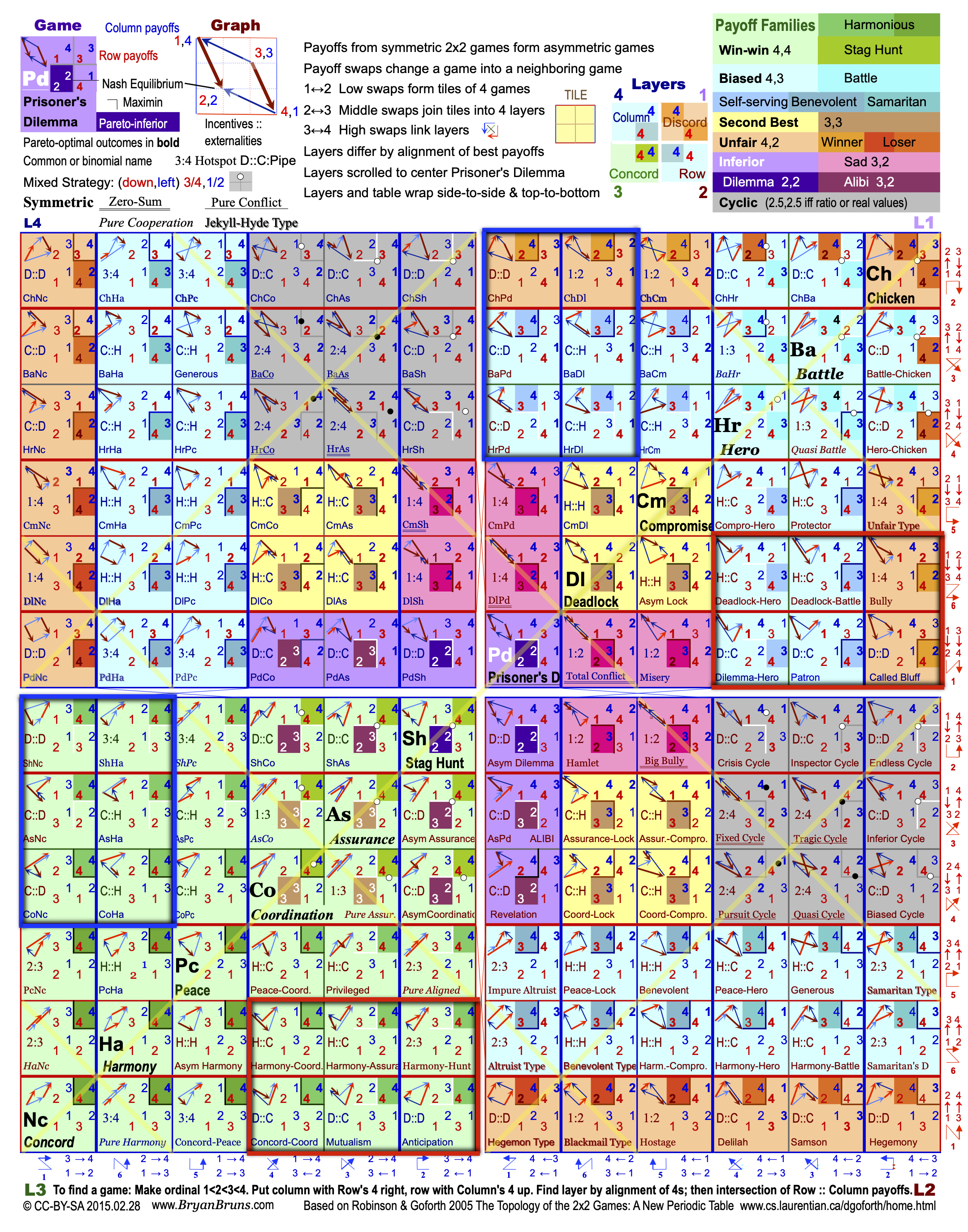}
\caption{\textbf{2x2 games susceptible to a one-player extortion}. To indicate the class of games that can be exploited, we use the periodic chart of 2x2 games \cite{robinson:2005:book,bruns:2015:atlas}. According to corollary \ref{def:susceptible_games} along with propositions \ref{def:one-player_restriction_I} and \ref{def:one-player_restriction_II}, the games outlined with a bold red border indicate cases where the row player is susceptible to extortion, while those with a bold blue border indicate susceptibility for the column player. To simplify the analysis and in accordance with assumption \ref{assump:uniqueNash_base}, we focus on games with a unique pure strategy Nash equilibrium, excluding (anti-)coordination and cyclic games. Under these constraints, $16,\overline{6}\%$ of 2x2 games are susceptible to the one-player extortion scheme.}
\label{fig:oneplayer_susceptible}
\end{figure}

In a one-player extortion without an external agent, the 2x2 games identified in figure \ref{fig:oneplayer_susceptible} remain susceptible to extortion. However, in this scenario, susceptibility depends not only on the game's structural properties but also on the magnitude of payoffs, as established in \ref{def:worthwhile_games}. This additional constraint further limits the set of games vulnerable to extortion.
Interestingly, we find that the win-win games vulnerable to extortion pose a higher risk for the extortioner when issuing the binding threat. If the extorted player decides to Decline, the extortioner player, $i_e$, could end up with a worse payoff than in the base game: ${u_{i_e}(s^{*A})>u_{i_e}(s^{*})>u_{i_e}(s^{*D})}$.

\subsubsection{Example}

To illustrate this scenario, we once again consider an asymmetric Concord base game that meets the required criteria


\begin{table}[!h]
    \centering
    \caption*{Base game ($G_B$)}
    \begin{tabular}{c|c|c|}
    &L&R \\
    \hline
    T& 4,6&8,4\\
    \hline
    B& 6,8& \textcolor{violet}{10,10} \\
    \hline
    \end{tabular}
    \label{tab:basegame}
\end{table}

The column player presents the following binding threat to the row player: \textit{``Either give me 2 (Euros), or if you decline, I give you 3 whenever you play Top."}

\begin{table}[!h]
\caption*{Manipulated game ($\Gamma_M$)}
\begin{subtable}[c]{0.23\textwidth}
\centering
\begin{tabular}{c|c|c|}
    &L&R \\
    \hline
    T& 2,8&6,6\\
    \hline
    B& 4,10& \textcolor{violet}{8,12} \\
    \hline
\end{tabular}
\subcaption*{Accept game ($G_A$)}
\end{subtable}
\begin{subtable}[c]{0.23\textwidth}
\centering
\begin{tabular}{c|c|c|}
    &L&R \\
    \hline
    T& \textcolor{violet}{7,3}&11,1\\
    \hline
    B& 6,8& 10,10 \\
    \hline
\end{tabular}
\subcaption*{Decline game ($G_D$)}
\end{subtable}
\end{table}

According to corollary \ref{def:max_profit} and the related propositions \ref{def:one-player_maxprofit} and \ref{def:min_profit}, the allowed range for the extractable profit is: $c_1^{max}=3$ and $c_1^{min}=0$, with $c_1^{max} > c_1 > c_1^{min}$. Similarly, corollary \ref{def:min_pay} and proposition \ref{def:one-player_maxpayment} determine the allowed range for the payment: $c_2^{max}=4$ and $c_2^{min}=2$, where $ c_2^{max} > c_2 > c_2^{min}$.

\section{Applications}

According to section \ref{sec:binding_threat}, the extortion scheme relies on strategic interactions being public, enabling outcome-contingent commitments to be made by an external agent or by the players before the game is played. This requirement is often met in real-world settings, meaning that numerous strategic interactions between individuals, organizations, and nation-states may be alarmingly susceptible to extortion \cite{ellsberg:1959:lecture, schelling:1960:book}. In modern society, multi-agents systems, where artificial agents engage in real-time strategic interactions with both humans and each other, are also susceptible, especially when deployed in permissionless blockchain environments. This vulnerability arises because blockchains function as decentralized, public ledgers where interactions are visible to all network users \cite{buterin:2013:whitepaper,kosba:2016:IEEESP,dipiero:2017:IEEECSE}. Moreover, blockchains enable smart contracts —self-executing programs that enforce predetermined conditions— making it possible to implement outcome-contingent commitments \cite{tennenholtz:2004:GEB, lavictoire:2014:AAAI, luu:2016:ACMCCS, oesterheld:2019:TaD}. By leveraging blockchain data, agents can detect profitable opportunities and use bots to automatically submit outcome-contingent threats, resembling MEV exploits \cite{mccorry:2019:FCDS, daian:2020:IEEESP,sun:2023:arXiv}.

\section{Discussion}
\label{sec:discussion}
Our study explores extortion through outcome-contingent payments, focusing on the counterintuitive scheme where an extortioner extracts profit by offering a reward instead of threatening punishment. By introducing the outcome-contingent commitment, the extortioner introduces sequentiality into a simultaneous-move game, thereby reshaping the game's structure. We derive the conditions under which this form of extortion is successful and identify the class of games susceptible to the extortion scheme. We find that games where a player achieves their best possible outcome at equilibrium are particularly vulnerable to extortion. This vulnerability arises when the player is willing to pay a fee to the extortioner to preserve the equilibrium that leads to their optimal outcome. Consequently, win-win (no-conflict) games are most prone to extortion. This insight has significant implications for economic and political contexts, where efficient equilibria are essential for maximizing welfare and fostering cooperative interactions. Our findings highlight the need to safeguard strategic interactions with desirable equilibria from extortionary tactics.

Our framework also contrasts non-coercive extortion with its coercive counterpart, where extortion relies on punishment ($c_2<0$) rather than reward ($c_2>0$). In comparison to non-coercive extortion, coercive extortion does not require outcome-contingent commitments, as outcome-independent punishments can be just as effective. In this case, as long as the fee to be extracted ($c_1$) is smaller than the potential punishment ($c_2$), the extortion succeeds. This contrast further underscores the counterintuitive nature of non-coercive extortion, where players can be manipulated into worse outcomes not through threats of punishment, but through incentives. This insight reveals how payments, widely considered as desirable, can paradoxically act as strategic threats.

We further analyse two variations of the extortion scheme. First, we consider the scenario where an external agent serves as the extortioner. This situation is closely aligned with mechanism design and implementation theory \cite{jackson:2001:SCW}, where an external entity designs the game to achieve specific objectives. Our scheme is particularly related to \textit{k-implementation}, in which an external designer influences game outcomes through non-negative monetary payments for different strategy profiles \cite{monderer:2003:ACMecom}.
Second, we explore the case where the extortioner is a player within the game, connecting our analysis to the literature on side-payments and transferable utilities \cite{jackson:2005:RES}. This setup is more restrictive than the scenario involving an external agent, as the extortioner's payoffs are now interdependent with the extorted player, introducing additional constraints on its feasibility.

The limitations of our proposed extortion scheme stem from the fundamental assumptions of game theory, where players are assumed to be self-interested and operate under common knowledge of rationality. If these assumptions are relaxed —such as considering players with bounded rationality, social preferences, or incomplete information— the outcomes of the extortion scheme may differ, potentially altering the feasibility of successful extortion.

Moreover, our study only focuses on pure strategies, as the credibility of commitments based on mixed strategies remains a topic of ongoing debate. Convincing individuals of commitments based on a randomization device is inherently challenging \cite{harrenstein:2007:AMAS,renou:2009:GEB}. 
Additionally, mixed strategies introduce strategic uncertainty, which may be undesirable for an extortioner seeking to ensure a profit through an outcome-contingent threat on a one-shot game \cite{vanDamme:1991:book,moore:2025:workingpaper}. However, it is important to note that the formulation in section \ref{sec:formulation} and theorem \ref{def:theorem_1} can be directly extended to include mixed strategies.

Further extensions can include allowing players to refuse payments \cite{bono:2014:EPE}, introducing multiple extortioners into the framework, and enabling players to counteract extortion attempts through counteroffers \cite{ramirez:2023:arXiv}. These aspects relate to the broader literature on commitment interactions, particularly in settings where commitments are made in reference to other commitments \cite{harrenstein:2007:AMAS}. Interestingly, a folk theorem can be derived for such interactions \cite{katz:1991:RAND}, much like in repeated games where extortion is also possible \cite{press:2012:PNAS, hilbe:2013:PNAS}. This parallel suggests that the presence of a folk theorem may signal the emergence of extortion opportunities. Exploring these connections in greater depth could provide valuable insights into both the persistence and mitigation of extortion in strategic interactions.


\subsection*{Acknowledgements}
We thank Chaitanya S. Gokhale, Iyad Rahwan and Arne Traulsen for valuable feedback and discussions.

\printbibliography

\onecolumn

\end{document}